\documentclass[reprint,aps,prl,showpacs,preprintnumbers,amsmath,amssymb,superscriptaddress]{revtex4-1}

\usepackage{graphicx}
\usepackage{dcolumn}
\usepackage{bm}
\usepackage{wasysym}
\usepackage{xspace}
\usepackage{amssymb}
\usepackage[mediumqspace, thinspace, amssymb]{SIunits}
\usepackage{color}

\newcommand{\dd}{\textrm{d}}

\newcommand{\be}{\begin{equation}}
\newcommand{\ee}{\end{equation}}

\newcommand{\fst}{\quad\textrm{.}}%
\newcommand{\com}{\quad\textrm{,}}%

\newcommand{\BFA}{BaFe$_2$As$_2$\xspace}
\newcommand{\Aig}{A$_\mathrm{1g}$\xspace}
\newcommand{\1}{(1~0~5)\xspace}
\newcommand{\2}{(2~0~6)\xspace}

\begin{document}

\title{Ultrafast structural dynamics of the Fe-pnictide parent compound \BFA}

\author{L. Rettig}
\email[Corresponding author: ]{laurenz.rettig@psi.ch}
\affiliation{Swiss Light Source, Paul Scherrer Institut, CH-5232 Villigen PSI, Switzerland}
\author{S. O. Mariager}
\affiliation{Swiss Light Source, Paul Scherrer Institut, CH-5232 Villigen PSI, Switzerland}
\author{A. Ferrer}
\affiliation{Institute for Quantum Electronics, Physics Department, ETH Z\"urich, CH-8093 Z\"urich, Switzerland}
\affiliation{Swiss Light Source, Paul Scherrer Institut, CH-5232 Villigen PSI, Switzerland}
\author{S. Gr\"ubel}
\affiliation{Swiss Light Source, Paul Scherrer Institut, CH-5232 Villigen PSI, Switzerland}
\author{J. A. Johnson}
\affiliation{Swiss Light Source, Paul Scherrer Institut, CH-5232 Villigen PSI, Switzerland}
\author{J. Rittmann}
\affiliation{Ecole Polytechnique F\'ed\'erale de Lausanne, Laboratoire de Spectroscopie Ultrarapide, ISIC, FSB, CH-1015 Lausanne, Switzerland}
\affiliation{Swiss Light Source, Paul Scherrer Institut, CH-5232 Villigen PSI, Switzerland}
\author{T. Wolf}
\affiliation{Karlsruhe Institute of Technology, Institut f\"ur Festk\"orperphysik, D-76021 Karlsruhe, Germany}
\author{S. L. Johnson}
\affiliation{Institute for Quantum Electronics, Physics Department, ETH Z\"urich, CH-8093 Z\"urich, Switzerland}
\author{G. Ingold}
\affiliation{Swiss Light Source, Paul Scherrer Institut, CH-5232 Villigen PSI, Switzerland}
\affiliation{SwissFEL, Paul Scherrer Institut, CH-5232 Villigen PSI, Switzerland}
\author{P. Beaud}
\affiliation{Swiss Light Source, Paul Scherrer Institut, CH-5232 Villigen PSI, Switzerland}
\affiliation{SwissFEL, Paul Scherrer Institut, CH-5232 Villigen PSI, Switzerland}
\author{U. Staub}
\affiliation{Swiss Light Source, Paul Scherrer Institut, CH-5232 Villigen PSI, Switzerland}

\date{\today}

\begin{abstract}
Using femtosecond time-resolved x-ray diffraction we investigate the structural dynamics of the coherently excited \Aig phonon mode in the Fe-pnictide parent compound \BFA. The fluence dependent intensity oscillations of two specific Bragg reflections with distinctly different sensitivity to the pnictogen height in the compound allow us to quantify the coherent modifications of the Fe-As tetrahedra, indicating a transient increase of the Fe magnetic moments. By a comparison with time-resolved photoemission data we derive the electron-phonon deformation potential for this particular mode. The value of ${\Delta \mu}/{\Delta z} = \unit{-(1.0-1.5)}{eV/\textrm{\AA}}$ is comparable with theoretical predictions and demonstrates the importance of this degree of freedom for the electron-phonon coupling in the Fe pnictides.

\end{abstract}


\maketitle

\begin{figure}[tb]
\includegraphics[width=8.6cm]{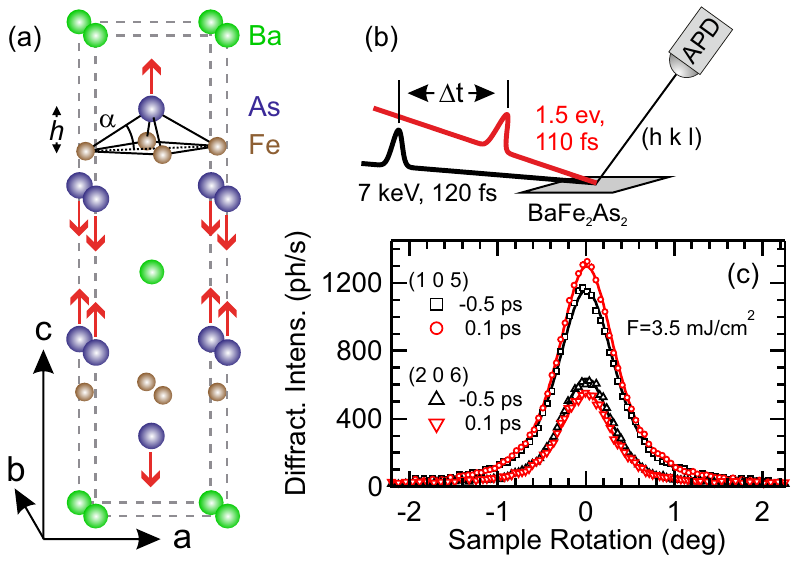}  
\caption{
\label{fig:fig1}
(color online) (a) Unit cell of \BFA with an Fe-As tetrahedron indicated. Red arrows show the atomic displacements of the \Aig phonon mode, which modulates the Fe-As tetrahedra angle $\alpha$. (b) Sketch of the experimental geometry. (c) Time-resolved rocking curves of the \1 and \2 reflection before (black) and at $\unit{100}{fs}$ after excitation (red).
}
\end{figure}

In the Fe pnictides, the complex interplay of the electronic, spin, orbital/ising-nematic and lattice degrees of freedoms leads to the emergence of a complex phase diagram, including structural transitions, spin-density wave (SDW) phases and high-temperature superconductivity. These phases emerge for electron and hole doping, but also for isovalent doping and external pressure~\cite{Mazin2010, Johnston2010}. The electronic structure and the magnetic properties depend very sensitively on the exact shape and size of the Fe-As tetrahedra, where an important degree of freedom is the pnictogen height $h$ above the Fe layers, which changes the Fe-As tetrahedra angle $\alpha$ (see Fig.~\ref{fig:fig1}(a)). A high sensitivity of the Fe magnetic moments on the pnictogen height with a rate of $\unit{6.8}{\mu_\mathrm{B}/\textrm{\AA}} $ has been predicted~\cite{Yin2008}, signifying a strong magneto-structural coupling~\cite{Yildirim2009, Egami2010, Johnston2010}. Similarly, an increase of the electron-phonon (e-ph) coupling strength was been found in calculations including magnetic ordering~\cite{Yndurain2009, Boeri2010}. Indeed, a universal relation between the Fe-As tetrahedra angle and the superconducting critical temperature $T_c$ has been proposed for the Fe pnictides~\cite{Lee2008, Johnston2010}, underlining the importance of structural degrees of freedom in these compounds.

The role of e-ph coupling for the mechanism of superconductivity in the Fe pnictides is still controversial. The average e-ph coupling constant has been found to be relatively weak both experimentally~\cite{Stojchevska2010, Mansart2010, Rettig2013} and theoretically~\cite{Boeri2008, Boeri2010} with $\lambda\lesssim 0.35$, insufficient to explain the high critical temperatures found in the pnictides in a conventional pairing scheme. However, due to the strong interplay of structural and magnetic degrees of freedom, a few phonon modes with enhanced magneto-structural coupling could still play an important role in the superconducting pairing mechanism~\cite{Egami2010, Yndurain2011}.

One such mode is the Raman active \Aig mode at the zone center corresponding to a displacement of the As ions perpendicular to the Fe layers with $\hbar \omega \approx \unit{22}{meV}$ (Fig.~\ref{fig:fig1}(a)). This mode directly modulates the pnictogen height $h$ and thereby the Fe-As tetrahedra angle $\alpha$. Coherent excitation of this mode has recently been observed using femtosecond (fs) time-resolved optical spectroscopy~\cite{Mansart2009}, and time-resolved THz spectroscopy demonstrated a transient resurrection of the magnetically ordered state during the coherent oscillations~\cite{Kim2012}. In addition, using time- and angle-resolved photoemission spectroscopy, a strong modulation of the chemical potential by the coherent \Aig mode has been observed~\cite{Avigo2013, Yang2014}. These observations demonstrate a significant coupling of the \Aig mode to both the electronic and the spin system, which is also the subject of recent theoretical investigations~\cite{Yndurain2011, Garcia-Martinez2013, Lee2014}.

So far no quantitative information about the ultrafast structural motions of the Fe-As lattice has been available. Structural information is crucial for determining the coupling of the various degrees of freedom to this coherent mode, and previous estimates of the structural displacement amplitudes relied solely on theoretical considerations. As such, even the direction of the displacive excitation towards smaller or larger tetrahedra angle $\alpha$ is controversial in the literature. Some papers argue that the excitation leads to a larger Fe-As distance and an increase of $\alpha$~\cite{Kim2012, Yang2014}, while a recent theoretical investigation predicts a transient resurrection of the SDW state for a decrease in $\alpha$~\cite{Lee2014}.

In this letter, we investigate the structural dynamics of the coherently excited \Aig mode in the Fe-pnictide parent compound \BFA using fs time-resolved x-ray diffraction. We observe the fluence dependent intensity oscillations of two specific Bragg reflections with opposite sensitivity to the pnictogen height. Calculations of the structure factor allow us to quantify the coherent displacement and oscillation amplitudes, yielding a transient increase of the tetrahedra angle $\alpha$ as large as $0.7^\circ$, compatible with a transient increase of the Fe magnetic moment. In addition, by comparing our results to the transient chemical potential shift observed in trARPES, we are able to quantify the e-ph deformation potential and determine the e-ph coupling constant $\lambda_{A_\mathrm{1g}}$ for this particular mode.

High-quality single crystals of \BFA were grown by a self-flux method~\cite{Hardy2010}. Time-resolved x-ray diffraction experiments were performed at the hard x-ray FEMTO slicing facility at the Swiss Light Source~\cite{Beaud2007} in an asymmetric diffraction configuration~\cite{Johnson2010} (Fig.~\ref{fig:fig1}(b)). The (0~0~1) surface of the cleaved \BFA single crystal was kept at $T=\unit{140}{K}$ during the measurements using a cryogenic nitrogen blower. This is above the N\'eel temperature $T_N\approx\unit{134}{K}$ and avoids a splitting of the Bragg peaks due to the orthorhombic distortion below $T_N$. The sample was excited at a $\unit{1}{kHz}$ repetition rate by $\unit{1.55}{eV}$ laser pulses with a duration of $\sim\unit{110}{fs}$ full-width at half maximum (FWHM). The sliced x-ray pulses ($\sim\unit{120}{fs}$ FWHM) were incident on the sample at a grazing angle of $\alpha_i=0.43^\circ$, matching the x-ray penetration depth to the optical penetration depth of $\sim\unit{25}{nm}$ at the x-ray energy of $\unit{7}{keV}$ ($\unit{260}{photons/pulse}$ at $\unit{2}{kHz}$). The x-ray beam was focused to $\unit{10}{\mu m}$ vertically and $\unit{300}{\mu m}$ horizontally to ensure a homogenous excitation of the probed volume. The overall time resolution was $\sim\unit{160}{fs}$. The diffracted x-ray photons were detected by a fast avalanche photodiode (APD).

Time-resolved rocking curves for a rotation about the sample surface normal for the \1 and \2 lattice reflections are shown in Fig.~\ref{fig:fig1}(c) before (black) and at $t=\unit{0.1}{ps}$ after excitation (red). The rocking curves are well described by a squared Lorentzian function (lines), which demonstrates the homogeneity and high crystal quality of the sample. The two reflections show a distinctly different response to the pump pulse: Whereas the \1 reflection shows a significant increase after excitation, the \2 reflection is decreased by the pump pulse. This indicates an ultrafast increase of the Fe-As tetrahedra angle $\alpha$ due to the displacive excitation of the coherent \Aig mode (see below). Both peak position and width remain constant for both peaks for up to $\sim\unit{1}{ps}$, indicating little influence of strain for these early times~\footnote{For $t>\unit{2}{ps}$, a strain wave is observed, which shifts the peak position and width. Therefore, we limit the analysis to early times.}.

\begin{figure}[tb]
\includegraphics[width=8.6cm]{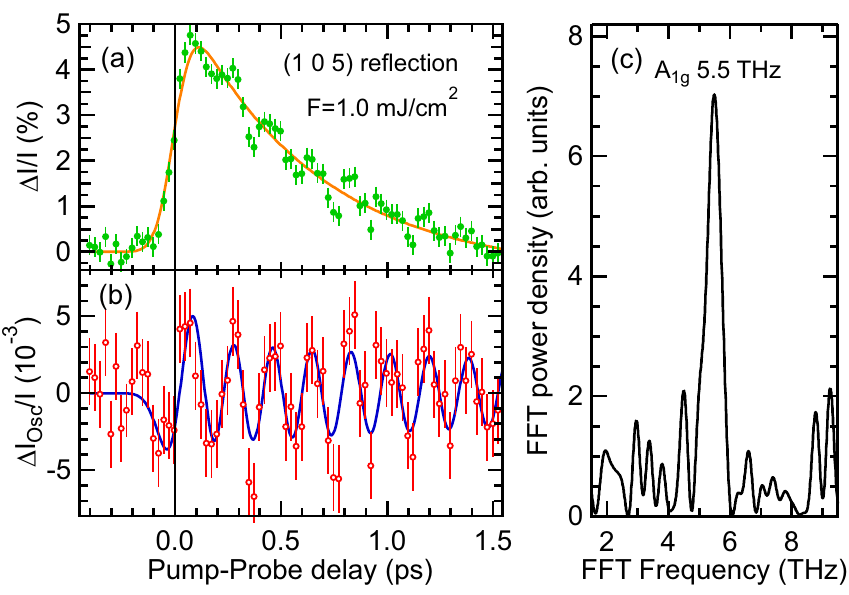}  
\caption{
\label{fig:fig2}
(color online) (a) Pump-induced change of the peak intensity of the \1 reflection as a function of pump-probe delay for an absorbed fluence of $F=\unit{1.0}{mJ/cm^2}$. (b) Oscillatory component after subtracting an exponential fit to the data (orange line in (a)). The blue line is a fit of a damped oscillator. (c) FFT spectrum of the data in (b), showing a clear peak at $\unit{5.5}{THz}$.
}
\end{figure}

The normalized pump-induced intensity change $\Delta I/I$ of the \1 reflection is shown in Fig.~\ref{fig:fig2}(a) as a function of pump-probe delay. After the ultrafast increase of the diffraction signal by $\sim5\%$ at $t_0$ within the experimental time resolution, the signal shows an exponential decay with a sub-ps timescale, superimposed by an oscillatory component. Such a behavior is typical for the displacive excitation of a coherent phonon. An abrupt change of the atomic potential landscape induced by the pump laser results in a new position of the potential minimum and leads to coherent oscillations around this new minimum~\cite{Zeiger1992}. The exponential decay of the intensity change reflects the relaxation of the excited energy potential surface, as well as the influence of lattice heating, which influences the x-ray signal via the Debye-Waller effect.

The oscillatory component is extracted from the data by subtracting an exponential function (orange line), and is shown in Fig.~\ref{fig:fig2}(b). A fit of a damped cosine function to the data yields a frequency of $f=\unit{5.46(9)}{THz}$, which is corroborated by the clear peak in the Fast Fourier Transformation (FFT) of the data shown in Fig.~\ref{fig:fig2}(c), centered at $\unit{5.5}{THz}$. This value is in very good agreement with the frequency of the coherent \Aig mode observed by time-resolved optical~\cite{Mansart2009}, THz~\cite{Kim2012} and photoemission~\cite{Rettig2012, Avigo2013, Yang2014}  experiments. The displacive nature of the excitation is supported by the phase $\Phi=\unit{-1.1(2)}{\pi}$ of the oscillation, which is very close to a pure cosine-like excitation, characteristic for the displacive excitation of coherent phonons (DECP)~\cite{Zeiger1992}.

\begin{figure}[tb]
\includegraphics[width=8.6cm]{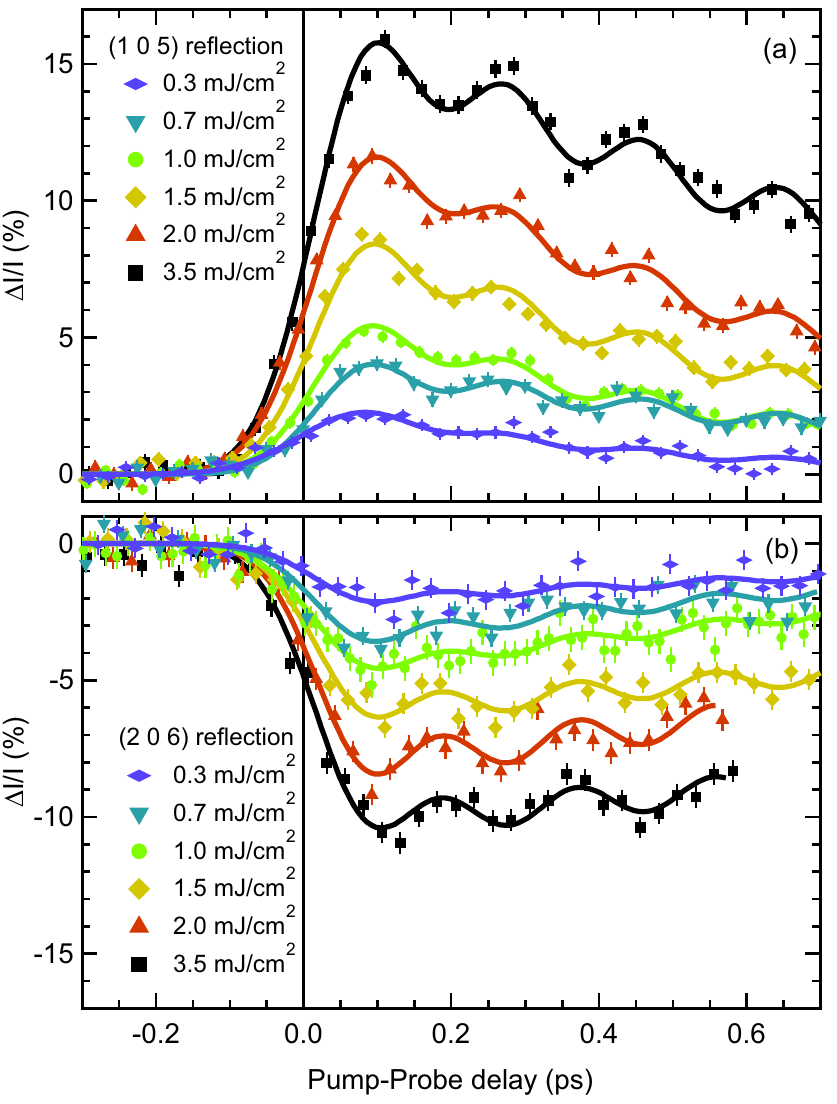}  
\caption{
\label{fig:fig3}
(color online) Pump-induced change of the peak intensity for (a) the \1 reflection and (b) the \2 reflection for various absorbed pump fluences. Lines are fits to equation \eqref{eqn:DECPfit}.
}
\end{figure}

For a more quantitative analysis of the coherent phonon oscillations, the pump-induced intensity change is shown for various fluences as a function of pump-probe delay for the \1 and \2 reflections in Fig.~\ref{fig:fig3}(a) and (b), respectively. In agreement with our observation in Fig.~\ref{fig:fig1}(c), the two reflections show an opposite behavior after excitation,  with a displacive increase in intensity for \1 and a decrease for \2, superimposed by the coherent oscillations of the \Aig mode. 

In order to determine the amplitudes of displacement and oscillations, we model the transient change in diffraction signal by the following expression for a displacively excited mode, consisting of a displacive and an oscillatory component:
\begin{multline}
\Delta I/I(t) = A_\mathrm{Disp}e^{(-t/\tau_\mathrm{Disp})} \\
+ A_\mathrm{Osc}\cos{(\omega t + \Phi)}e^{(-t/\tau_\mathrm{Osc})}\fst
\label{eqn:DECPfit}
\end{multline}
Here, $A_\mathrm{Disp}$ and $A_\mathrm{Osc}$ are the amplitudes of displacive and oscillatory components, $\tau_\mathrm{Disp}$ and $\tau_\mathrm{Osc}$ the relaxation timescales of the displacive and oscillatory part, and $\omega$ and $\Phi$ the frequency and phase of the oscillation, respectively. In order to take the finite temporal resolution of the experiment into account, equation~\eqref{eqn:DECPfit} is convolved by a Gaussian profile with $\unit{160}{fs}$ FWHM. Due to the limited time window of the data in Fig.~\ref{fig:fig3}, the frequency and phase of the oscillations have been fixed at the values determined from Fig.~\ref{fig:fig2}.

\begin{figure}[tb]
\includegraphics[width=8.6cm]{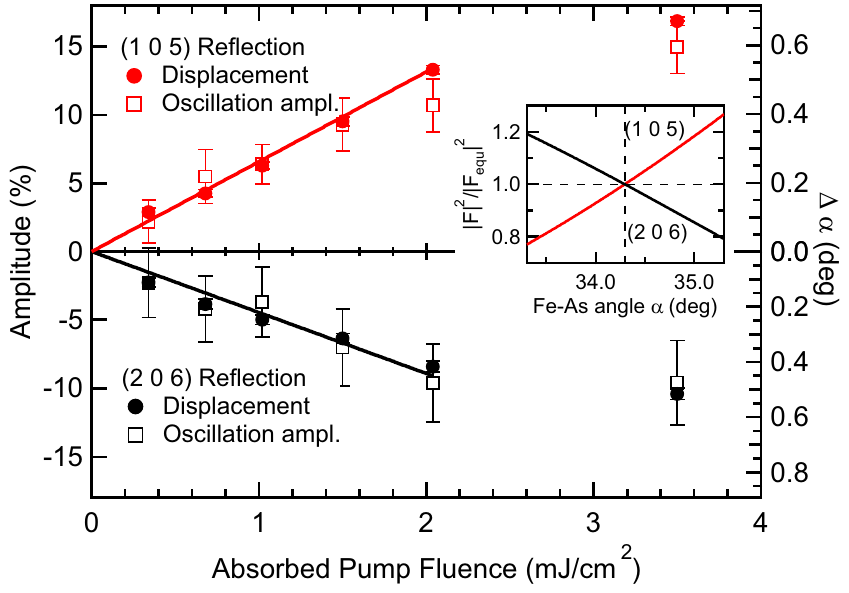}  
\caption{
\label{fig:fig4}
(color online) Displacement amplitudes $A_\mathrm{Disp}$ (circles) and oscillation amplitudes $A_\mathrm{Osc}$ (squares) for the \1 (red) and \2 reflection (black), derived from the fits in Fig.~\ref{fig:fig3}. Error bars are 95$\%$ confidence intervals, and lines are linear fits. Inset: Calculated normalized diffraction intensity of the \1 (red) and \2 reflection (black) as a function of Fe-As tetrahedra angle $\alpha$. The vertical line marks the equilibrium position.
}
\end{figure}

Fits of the data to equation~\eqref{eqn:DECPfit} are shown in Fig.~\ref{fig:fig3} as lines and reproduce the data very nicely. Figure~\ref{fig:fig4}(a) summarizes the amplitudes for the displacive and oscillatory components $A_\mathrm{Disp}$ and $A_\mathrm{Osc}$ as solid and open symbols, and for the \1 (red) and \2 (black) reflection, respectively. Both components show a linear behavior for low fluences ($F<\unit{2}{mJ/cm^2}$), and saturate for the largest fluences. Remarkably, the amplitudes of the oscillatory and displacive components agree very well within the accuracy of the experiment, as expected from the DECP model of coherent phonon excitation~\cite{Zeiger1992,Huber2014}. The larger error bars for $A_\mathrm{Osc}$ are due to the deconvolution with the limited time resolution, which suppresses the experimentally observed oscillation amplitudes. The saturation behavior observed for the highest pump fluences indicates a nonlinear response of the excited atomic potential to the electronic excitation level or anharmonicities in the atomic potential~\cite{DeCamp2001, Fritz2007}.

In order to relate the amplitudes to the microscopic atomic displacements and the modification of the Fe-As tetrahedra, we calculate the diffraction intensities of the \1 and \2 reflections as a function of the Fe-As tetrahedra angle $\alpha$ in the kinematic approximation~\cite{SOM}. The calculated  normalized intensities are shown in the inset of Fig.~\ref{fig:fig4} and nicely reproduce the opposite slope of the two reflections with respect to $\alpha$. We find a relative intensity change of $\unit{+24.8}{\%/deg}$ for the \1 reflection, and $\unit{-20.2}{\%/deg}$ for the \2 reflection. Comparison with the experimental amplitudes allows us to quantitatively determine the average atomic displacement induced by the coherent \Aig mode (right axis of Fig.~\ref{fig:fig4}). For the highest fluence of $F=\unit{3.5}{mJ/cm^2}$, we find displacement and oscillation amplitudes as large as $\Delta \alpha = 0.68^\circ$, corresponding to an initial displacement of the As atoms by $\Delta h = \unit{7}{pm}$, more than $5\%$ of the equilibrium pnictogen height. A linear fit at low fluences yields an amplitude of $\unit{0.26(1)}{deg/mJcm^{-2}}$ ($\unit{1.32(6)}{pm/mJcm^{-2}}$) for the \1 reflection, and $\unit{0.22(2)}{deg/mJcm^{-2}}$ ($\unit{1.12(11)}{pm/mJcm^{-2}}$) for the \2 reflection. The slight difference of the values found for the two reflections could be due to the limits of the kinematic approximation, or indicate further structural components in the coherent oscillations beyond the vertical motion of the As coordinate~\cite{Avigo2013}. 

The quantification of the absolute oscillation amplitude of the coherent \Aig mode offers important insight into the interplay of structural and electronic degrees of freedom, by comparison to the imprint of the same coherent phonon oscillation on the electronic structure observed by time-resolved photoemission spectroscopy. Yang et \emph{al.}~\cite{Yang2014} report a coherent modulation of the chemical potential of $\sim\unit{40-50}{meV/mJcm^{-2}}$. Comparing this with the structural information obtained here, and taking the different probe depth of the x-rays and the photoemission process into account~\cite{SOM}, we estimate a deformation potential of the As \Aig mode of ${\Delta \mu}/{\Delta \alpha} = \unit{-(0.05-0.08)}{eV/deg}$ or ${\Delta \mu}/{\Delta z} = \unit{-(1.0-1.5)}{eV/\textrm{\AA}}$. This value is in reasonable agreement with the deformation potentials predicted by density functional theory for variation of the As heights in various FeAs compounds~\cite{Singh2008, Yndurain2009, Yndurain2011, Dhaka2013, Lee2014}, which are on the order of $\dd E/\dd z \sim \unit{1.5-2}{eV/\textrm{\AA}}$. The deformation potential is also comparable to the deformation potential of the coherent \Aig mode in Bismuth~\cite{Papalazarou2012}. 

The e-ph deformation potential allows us to determine the e-ph coupling constant for the \Aig mode, which is related to the electronic deformation potential as~\cite{Yndurain2011}:
\be
\lambda_{A_\mathrm{1g}}=\frac{1}{4 M_\mathrm{As}\omega_{A_\mathrm{1g}}^2} N(E_F)\left(\frac{\dd E}{\dd z}\right)^2\com
\label{eqn:lambda}
\ee
where $M_\mathrm{As}$ is the As atomic mass, $\omega_{A_\mathrm{1g}}$ the \Aig phonon frequency and $N(E_F)$ the electronic density of states at the Fermi level. Using $N(E_F)\sim\unit{2-5}{states/eV/\textrm{f.u.}}$ from density functional theory~\cite{Singh2008}, we obtain $\lambda_{A_\mathrm{1g}}=0.05-0.30$. This value for $\lambda$ is in the same range as the total e-ph coupling constant $\lambda_\mathrm{tot}$ in \BFA, as predicted theoretically~\cite{Boeri2010}, and measured by time-resolved optics and photoemission~\cite{Mansart2010, Rettig2013}. The total e-ph coupling is the sum of the coupling to all modes in the system. Thus, our finding of $\lambda_{A_\mathrm{1g}}\sim\lambda_\mathrm{tot}$ indicates that the \Aig mode plays a major role for the e-ph coupling in \BFA. 

Finally, the observation of a displacive excitation towards \emph{larger} pnictogen height and \emph{larger} tetrahedra angles $\alpha$ implies an ultrafast increase of the Fe magnetic moments~\cite{Yin2008, Yildirim2009, Egami2010, Johnston2010}. This enhanced magnetic moments could in principle be responsible for the ultrafast resurrection of the SDW order~\cite{Kim2012}. However, the atomic displacement at the fluence of $F\approx\unit{0.5}{mJ/cm^2}$ based on band structure calculations~\cite{Yin2008} yields a maximal increase of the magnetic moment of $~0.08\mu_B$, which corresponds to a relative change of $\sim11\%$~\cite{Avci2012}. This leads to an increase of T$_N$ by $\sim\unit{20}{K}$, if we consider the ratio of the SDW transition temperature and the Fe magnetic moments of $\unit{200-250}{K/\mu_B}$ observed in Ba$_{1-x}$K$_x$Fe$_2$As$_2$~\cite{Avci2012} and BaFe$_2$(As$_{1-x}$P$_x$)$_2$~\cite{Allred2014}. Given the fact that the transient magnetic ordering has also been observed $>\unit{100}{K}$ above $T_N$~\cite{Kim2012}, the increased magnetic moments cannot be solely responsible for the resurrection of SDW order, but the transiently modified nesting properties of the band structure need to be also considered~\cite{Kim2012}.

In conclusion, using time-resolved x-ray diffraction, we investigated the coherent structural dynamics of the displacively excited \Aig phonon mode oscillations in \BFA. Based on structure factor calculations we determine the absolute oscillation and displacement amplitudes, and the direction of the modulation. Comparison of the absolute oscillation amplitude of the As coordinate to time-resolved photoemission experiments allows us to determine the \Aig deformation potential ${\Delta \mu}/{\Delta z} = \unit{-(1.0-1.5)}{eV/\textrm{\AA}}$. An estimate of the e-ph coupling constant underlines the importance of this structural degree of freedom in the Fe pnictides. Our findings are consistent with a transient increase of the Fe magnetic moment, which however is predicted to be too small to explain the observed transient resurrection of the SDW phase in \BFA.

\begin{acknowledgments}
Time-resolved x-ray diffraction experiments were performed on the X05LA beamline at the Swiss Light Source, Paul Scherrer Institut, Villigen, Switzerland. We thank I. Eremin for stimulating discussions, and J. Fink for support in sample supply. We acknowledge support in sample characterization from C. Bernhard and M.A. Uribe-Laverde, and we thank D. Grolimund for experimental support. We acknowledge financial support by the NCCR Molecular Ultrafast Science and Technology (NCCR MUST), a research instrument of the Swiss National Science Foundation (SNSF).

\end{acknowledgments}

%

\end{document}